\documentclass{kapproc} 
\usepackage{graphicx}
\usepackage{epsfig}

\def\lesssim{\mathrel{\hbox{\rlap{\hbox{\lower4pt\hbox{$\sim$}}}\hbox{$<$}}}}
\def\gtrsim{\mathrel{\hbox{\rlap{\hbox{\lower4pt\hbox{$\sim$}}}\hbox{$>$}}}}

\kluwerbib

\def\lesssim{\mathrel{\hbox{\rlap{\hbox{\lower4pt\hbox{$\sim$}}}\hbox{$<$}}}}
\def\gtrsim{\mathrel{\hbox{\rlap{\hbox{\lower5pt\hbox{$\sim$}}}\hbox{$>$}}}}

\newcommand{\begen}{\begin{enumerate}}
\newcommand{\enen}{\end{enumerate}}

\newcommand{\lumnue}{L_{\nu_e}}

\newcommand{\lumanue}{L_{\bar{\nu}_e}}

\newcommand{\avenue}{\langle\varepsilon_{\nu_e}\rangle}
\newcommand{\aveanue}{\langle\varepsilon_{\bar{\nu}_e}\rangle}
\newcommand{\aveunu}{\langle\varepsilon_{\nu_\mu}\rangle}

\newcommand{\begit}{\begin{itemize}}
\newcommand{\enit}{\end{itemize}}
\newcommand{\beq}{\begin{equation}} 
\newcommand{\eeq}{\end{equation}} 
\newcommand{\beqa}{\begin{eqnarray}} 
\newcommand{\eeqa}{\end{eqnarray}} 
\newcommand{\p}{\partial}    
\newcommand{\pr}{^\prime}

\begin{document}

\articletitle{Protoneutron Star Winds}

\author{Todd A. Thompson\footnote{Hubble Fellow}}
\affil{Astronomy Department and Theoretical Astrophysics Center, \\ 
601 Campbell Hall, The University of California, Berkeley, CA 94720}
\email{thomp@astro.berkeley.edu}

\begin{abstract}
Neutrino-driven winds are thought to accompany the Kelvin-Helmholtz
cooling phase of nascent protoneutron stars in the first seconds after
a core-collapse supernova.  These outflows are a likely candidate as 
the astrophysical site for rapid neutron-capture nucleosynthesis (the $r$-process).  
In this chapter we review the physics of protoneutron star winds
and assess their potential as a site for the production of the
heavy $r$-process nuclides.

We show that spherical transonic protoneutron star winds do not produce 
robust $r$-process nucleosynthesis for `canonical' neutron stars with 
gravitational masses of 1.4\,M$_\odot$ and coordinate radii of  10\,km.
We further speculate on and review some aspects of neutrino-driven winds from
protoneutron stars with strong magnetic fields.  
\end{abstract}


\pagebreak

\section{Introduction}

\subsection{$r$-Process Nucleosynthesis}

A complete and self-consistent theory of the origin of the 
elements has been the grand program of nuclear
astrophysics since the field's inception.  Of the several 
distinct nuclear processes which combine to produce the
myriad of stable nuclei and isotopes we observe, 
none has generated more speculation than 
$r$-process nucleosynthesis (Wallerstein 1997). 
The $r$-process, or rapid neutron-capture process, 
originally identified in Burbidge et al. (1957) and Cameron (1957), 
is a mechanism for nucleosynthesis by which seed nuclei 
capture neutrons on timescales much shorter than those for $\beta^-$ decay.
The rapid interaction of neutrons with heavy, neutron-rich, seed nuclei allows
a neutron capture-disintegration equilibrium to establish itself 
among the isotopes of each element.  The nuclear flow proceeds
well to the neutron-rich side of the valley of $\beta$-stability
and for sufficient neutron-to-seed ratio ($\gtrsim$100) the $r$-process
generates the heaviest nuclei (e.g., Eu, Dy, Th, and U), forming
characteristic abundance peaks at $A\sim80$, 130, and 195 
(e.g.~Burbidge et al.~1957; Meyer \& Brown 1997; Wallerstein et al.~1997).

The neutron-to-seed ratio is the critical parameter in determining 
if $r$-process nucleosynthesis succeeds in producing nuclei up
to and beyond the third abundance peak. If the neutron-to-seed ratio
is too small, the $r$-process may only nuclei up to the first or second
abundance peak. In a given
hydrodynamical flow, the neutron-to-seed ratio is set primarily by
the entropy ($s_{\rm a}$; `a' here stands for `asymptotic') of the flow, the neutron richness of the matter,  
and the dynamical timescale for expansion.  
The neutron richness is generally quantified by the 
the electron fraction ($Y_e^{\rm a}$, the number density of electrons per baryon).
The dynamical timescale ($\tau_{\rm dyn}$) is a characteristic time for expansion,
i.e.~the $e$-folding time for temperature or density at a given temperature.
Thus, $s_{\rm a}$, $Y^{\rm a}_e$, and $\tau_{\rm dyn}$ set the neutron-to-seed ratio.  The higher the entropy,
the lower the electron fraction, and the shorter the dynamical timescale, 
the larger the neutron-to-seed ratio and the higher in $A$ the
nucleosynthetic flow will proceed (e.g.~Hoffman, Woosley, \& Qian 1997; Meyer \& Brown 1997).

\subsection{Observational Motivation: A Remarkable Concordance}
\label{section:observe}

\begin{figure}[t]
\vspace*{-1.in}
\hspace*{-.5cm}
\includegraphics[height=15cm,width=12cm]{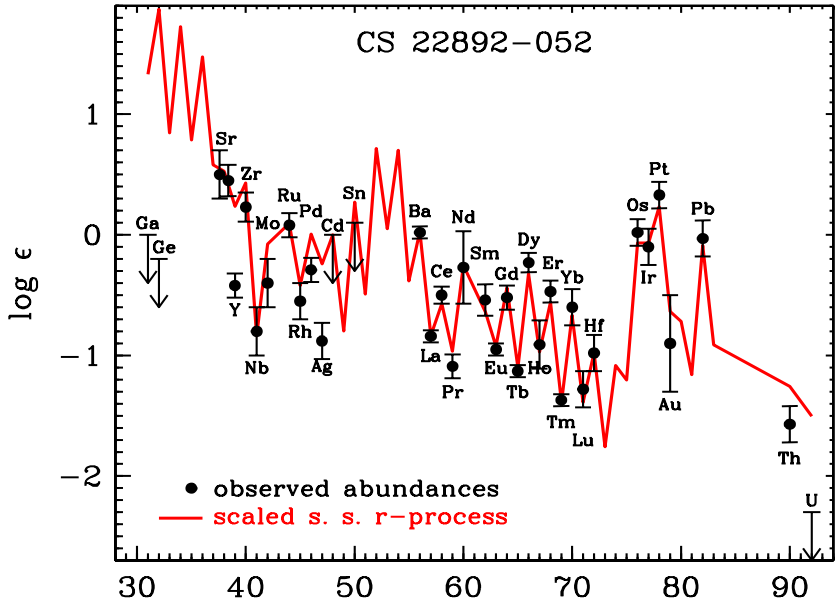}
\vspace*{-1.4in}
\caption{The scaled solar $r$-process abundances (solid line) and the 
abundances from CS 22892-052 (dots with error bars) as a function of 
atomic number (Chris Sneden, private communication).}
\label{fig:sneden}
\end{figure}

Recent observations of neutron-capture elements in ultra-metal-poor ([Fe/H]$\lesssim-2.5$] 
halo stars (Sneden et al.~1996; Burris et al.~2000; McWilliam et al.~1995a,b; 
Cowan et al.~1996; Westin et al.~2000; Hill et al.~2001) show remarkable agreement 
with the scaled solar $r$-process abundance pattern for $A\gtrsim135$.
Particularly for atomic numbers between $N=55$ and $N=75$, the distribution 
of abundances in these halo stars is identical with solar.  
Prototypical of this class are the stars CS 22892-052, BD +17$^{\rm o}$3248 and HD 115444
(Cowan \& Sneden 2002).  Figure \ref{fig:sneden} shows the
scaled solar $r$-process abundances (solid line) together with those
from the ultra-metal-poor halo star CS 22892-052 (points with error bars) as a function
of atomic number.  Note both the tight correspondence between the sun
and  CS 22892-052 above the second $r$-process peak and the large 
discrepancies at lower atomic number, in particular the elements Sn, Ag, and Y.
The fact that a class of very old halo stars share the same relative 
abundance of $r$-process elements above the second peak 
suggests a universal mechanism for producing these nuclei, which must act 
early in the chemical enrichment history of the galaxy. 

The fact that observations of ultra-metal-poor halo stars show significant
and scattered deviations from the scaled solar $r$-process abundance pattern
below $A\sim130$ has implied that there are perhaps two or more $r$-process
sites (Qian, Vogel, \& Wasserburg 1998; Wasserburg \& Qian 2000; Qian \& Wasserburg 2000).  
In these models of $r$-process enrichment, distinct astrophysical sites account for the
observed $r$-process abundances below and above $A\sim130$.  
Important in these considerations are the constraints set on the
site or sites of the $r$-process by the integrated galactic $r$-process budget.
For example, if all supernovae produce neutron stars that are accompanied
by protoneutron star winds that produce a robust $r$-process, then,
given a supernova rate of one every 50-100 years, 10$^{-5}-$10$^{-6}$\,M$_\odot$
of $r$-process material must be injected into the interstellar medium
per event (Qian 2000).  These numbers are to be contrasted with those from any other
potential $r$-process site.  For example, if the material possibly ejected
to infinity in the formation of a black hole from neutron-star/neutron-star
mergers accounts for the total galactic $r$-process budget
(Freiburghaus, Rosswog, \& Thielemann 1999, and references therein; Rosswog et al.~1999), 
then 10$^{-1}-$10$^{-2}$\,M$_\odot$ of $r$-process material must be ejected per 
event, given an event rate of one every 10$^5$ years (Kalogera et al.~2001; 
Qian 2000 and references therein).

Because the total $r$-process budget of the galaxy is fairly well known,
any potential site must do more than just match the entropy, electron fraction,
and dynamical timescales required for a high neutron-to-seed ratio. Simply
attaining the necessary physical conditions for a neutron-to-seed ratio above
$\sim$100 is not sufficient.  The astrophysical site, given an event rate,
must be able to consistently produce the robust $r$-process abundance pattern
in ultra-metal-poor halo stars in accordance with the total galactic $r$-process
budget.  Thus, if $M_{\rm ej}^{\rm r}$ is the total mass of $r$-process
material produced per event, the process of assessing an astrophysical site's potential 
for the $r$-process is merely a matter of mapping the region in 
$s_{\rm a}-Y_e^{\rm a}-\tau_{\rm dyn}-M_{\rm ej}^{\rm r}$ space physically accessible to 
the potential site.  In what follows here we review the results for just-post-supernova,
neutrino-driven protoneutron star winds.

\subsection{Protoneutron Star Winds}

The successful two-dimensional supernova explosion obtained
by Burrows, Hayes, \& Fryxell (1995) shows clearly a post-explosion
neutrino-driven wind emerging into the evacuated region above
the newly formed protoneutron star and behind the rapidly expanding
supernova shock.  Although no study was made of this outflow
as a function of progenitor, such a wind phase might naturally 
accompany the post-explosion cooling epoch in many core-collapse
supernovae.  

A multiple of $10^{53}$\,erg of binding energy 
will be lost as the nascent neutron star cools and contracts 
over its Kelvin-Helmholtz cooling time ($\sim10-20$ seconds).
This energy will be carried away predominantly by all species 
of neutrino. A small fraction of that energy will be deposited in the 
surface layers of the protoneutron star, ablating material
from its upper atmosphere and driving a hydrodynamical wind.  Whether 
this wind succeeds in escaping to infinity, or is prevented by 
fallback and reverse shocks as the supernova shock encounters the density 
stratifications of the overlying stellar mantle is an important future
area of study.  In addition, the actual emergence of the wind -
how it overcomes the non-zero pressure of the matter exterior to
the protoneutron star - has not been fully explored. 
In this work, however, we assume the existence of such outflows 
as evidenced by the calculation of Burrows, Hayes, \& Fryxell (1995)
as well as those of Janka \& M\"{u}ller (1995), and 
Takahashi, Witti, \& Janka (1994).  Our goal is to explore the basic physics
of neutrino-driven outflows and assess this site as a candidate for $r$-process
nucleosynthesis.

The basic scenario is as follows: as the supernova shock propagates
outward in a successful explosion, the pressure in the
region between the protoneutron star and the shock decreases and the wind,
powered by neutrino heating, emerges into the post-shock ejecta.
The surface of the protoneutron star is hot (temperatures of $\sim5$ MeV)
and has a low electron fraction (typically, $Y_e\sim0.1$). The matter
there is composed of relativistic charged leptons, free nucleons, and
trapped photons. As the wind is driven outward, the matter descends a
gradient in density and temperature.  The wind material is heated only
within the first $\sim$50\,km, primarily via the charged-current
absorption processes on free nucleons $\nu_e n\rightarrow p e^-$ and
$\bar{\nu}_e p \rightarrow n e^+$, and its entropy increases
concomitantly. As the temperature of the matter drops below $\sim1$ MeV,
nucleons combine into alpha particles, neutrino heating ceases and the
material expands adiabatically with entropy $s_{\rm a}$.  These
charged-current processes also set the asymptotic electron fraction.  As
the matter expands away from the protoneutron star surface and the
chemical equilibrium between $\nu_e$ and $\bar{\nu}_e$ neutrinos obtained
near the neutrino decoupling radius (the neutrinosphere)
is broken, the luminosity and energy density of the electron-type neutrino
species determine the electron fraction.  Typically, within 10\,km of the
protoneutron star surface $Y_e$ asymptotes to $Y_e^{\rm a}$. At a
temperature of $\sim$0.5 MeV (a radius in most models of $\sim$100\,km)
alpha particles combine with the remaining free neutrons in an
$\alpha$-process to form seed nuclei  (Woosley \& Hoffman 1992).
At this location in the temperature profile, it is
the steepness of the density or temperature gradient that sets
the dynamical timescale of the wind.  Simply stated, short dynamical
timescales are favored for the $r$-process because for faster expansions
there is less time to build seed nuclei, and, hence, the neutron-to-seed
ratio is preferentially larger, all else being equal.
When the wind material finally reaches a temperature of $\sim$0.1 MeV (at
a radius of several hundred kilometers) the $r$-process may begin if the
neutron-to-seed ratio, as set by $Y_e^{\rm a}$, $s_{\rm a}$, and
$\tau_{\rm dyn}$, is sufficiently high.

Because the success of the $r$-process is so dependent on the
neutron-to-seed ratio and, hence, the electron fraction, entropy, 
and dynamical timescale of the nucleosynthetic environment,
we focus on these quantities  and their sensitivity to various
parameters of the protoneutron star.  In addition to establishing 
correlations between these quantities in neutrino-driven wind environments, 
we must also fold in the constraint on the total mass ejected.  We will show 
in \S\ref{section:results} that this constraint on $M_{\rm ej}^{\rm r}$ significantly limits the 
$s_{\rm a}-Y_e^{\rm a}-\tau_{\rm dyn}$ space of relevance for winds.

\subsection{Previous Work}

Duncan, Shapiro, \& Wasserman (1986) were the first to explore
the physics of neutrino-driven winds.  They identified a number
of important scaling relations and the basic systematics
and dependencies of the problem.  In addition, they explored the
relative importance of the neutrino and photon luminosities in
determining the resulting hydrodynamics.
Recent investigations have focused on the potential of these outflows
for $r$-process nucleosynthesis as suggested in Woosley \& Hoffman (1992).
Witti, Janka, \& Takahashi (1994) and Takahashi, Witti, \& Janka (1994) 
showed that although interesting $\alpha$-process
nuclei were created in their models of protoneutron star winds, conditions
for a successful $r$-process fell short in entropy by a factor of $\sim5$.
Qian \& Woosley (1996) made analytical estimates of the fundamental wind 
properties and systematics and compared their scalings to numbers from hydrodynamical 
simulations.  They further explored interesting variations to their models such as
inserting artificial heating sources and applying an external boundary pressure.
These results were followed by nucleosynthetic calculations, taking wind trajectories
as the time-history of Lagrangean mass elements in Hoffman, Woosley, \& Qian (1997).
These efforts also showed that with reasonable variations in the input physics
protoneutron star winds do not achieve high enough entropy for the dynamical
timescales and electron fractions derived.  

Cardall \& Fuller (1997) extended the analytical work of Qian \& Woosley (1996)
to general-relativistic flows.  They found that significant enhancements
in entropy might be obtained from compact neutron stars with large $M/R$.
Several general-relativistic treatments of the problem followed, including
the work of Otsuki et al.~(2000), Sumiyoshi et al.~(2000), Wanajo et al.~(2001),
and Thompson, Burrows, \& Meyer (2001).  In the last of these, purely transonic 
steady-state wind solutions were derived in general relativity, with a simultaneous 
solution for the evolution of the electron fraction in radius and a careful
treatment of the boundary conditions.

Note that we distinguish here between models of protoneutron star {\it winds}
and {\it bubbles}.  The former is typified by those works already discussed,
in which the wind velocity approaches the local speed of sound, or a large
fraction of the speed of sound somewhere in the flow.  Such winds were 
realized in the self-consistent supernova calculations in Burrows, Hayes, \& Fryxell (1995).
In contrast, in the work of Woosley et al.~(1994) the protoneutron star 
outflow reached speeds of only a very small fraction of the sound speed in the 
region between the protoneutron star and the expanding supernova shock.
Approximately 18 seconds after collapse and explosion,
Woosley et al.~(1994) obtained entropies of $\sim$400 
(throughout, we quote entropy in units of k$_{\rm B}$ per baryon), 
long dynamical timescales, and electron fraction ($Y_e$) in the range $0.36-0.44$.  
However, in their model the supernova shock reached only 50,000 km at these late times.  In turn, 
this external boundary caused the wind material to move slowly. It remained in the 
heating region for an extended period, thus raising the entropy
above what any simulation or analytical calculation has since obtained.  
Although the $r$-process proceeded to the third abundance peak in their calculation, 
nuclei in the mass range near $A\sim90$ (particularly, $^{88}$Sr, $^{89}$Y, and $^{90}$Zr)
were overproduced by more than a factor of 100.


\subsection{This Review}

In \S\ref{section:hydro}, we review 
the fundamental and general equations for time-independent 
energy-deposition-driven winds in Newtonian gravity and in general relativity.
We also discuss the integrals of the flow and our numerical procedure for 
solving the relevant equations.
In \S\ref{section:pnswinds} we discuss some of the particulars of modeling
protoneutron star winds, including the neutrino heating function, the equation
of state, and the evolution of the electron fraction.  We further critically examine
several of the underlying assumptions that must accompany any such model.
\S\ref{section:results} summarizes our results for spherical winds.
In \S\ref{section:magnetic} we consider the possible effects of magnetic fields
and speculate on a number of issues in need of more thorough investigation.
In \S\ref{section:summary} we summarize and conclude.


\section{Hydrodynamics}
\label{section:hydro}

\subsection{The Newtonian Wind Equations}
\label{section:newtequations}

Assuming time-independent wind solutions, the equation for mass conservation is simply
\beq
{\bf \nabla\,\,\cdot}\,\,(\rho{\bf v})=0,
\label{con1} 
\eeq
implying that the mass-outflow rate ($\dot{M}$) of a wind is a constant in radius.
In spherical symmetry, $\dot{M}=4\pi r^2 \rho v={\rm constant}$.  This expression
yields a differential equation for the evolution of the matter velocity in radius,
\beq
\frac{1}{v}\frac{dv}{dr}=-\frac{1}{\rho}\frac{d\rho}{dr}-\frac{2}{r}.
\label{con}
\eeq
The equation for momentum conservation, neglecting the mass of the wind itself, is simply
\beq
v\frac{dv}{dr}=-\frac{1}{\rho}\frac{dP}{dr}-\frac{GM}{r^2}+F_\nu,
\label{mom}
\eeq
where $M$ is the total mass of the protoneutron star. Although we include it
here for completeness, the radiation force due to the neutrinos ($F_\nu$) can
be safely neglected.  This approximation is justified because
the neutrino Eddington luminosity ($L_\nu^{\rm Edd}=4\pi G M c/\kappa_\nu$)
is much larger than the neutrino luminosities that accompany the 
protoneutron star cooling/wind evolutionary phase.
$\kappa_\nu$ is the total neutrino opacity and is dominated
by $\nu_e n\rightarrow p e^-$ and $\bar{\nu}_e p\rightarrow n e^+$
for the electron and anti-electron neutrino, respectively.  The $\mu$- and $\tau$-neutrino
opacity is dominated by neutral-current scattering off free nucleons, the 
wind heating region being unpopulated by nuclei.  Including these processes,
one finds that $L_\nu^{\rm Edd}\gtrsim10^{55}$ erg s$^{-1}$.  We are thus
safe in taking $F_\nu=0$ because we consider winds with only 
$L^{\rm tot}_\nu<5\times10^{52}$ erg s$^{-1}$.

Because neutrinos contribute heating and cooling to the flow, we must couple 
to these equations to the first law of thermodynamics.  We define the net 
specific heating rate, $\dot{q}={\rm Heating}-{\rm Cooling}$,
so that 
\beq
\frac{d\epsilon}{dt}=\dot{q}=T\frac{ds}{dt}+\frac{P}{\rho^2}\frac{d\rho}{dt},
\eeq
where $d/dt=[\p/\p t+v\cdot\nabla]$.  In the steady state,
$\p/\p t=0$ and we obtain
\beq
\dot{q}=Tv\frac{ds}{dr}=C_V v \frac{dT}{dr}-\frac{vT}{\rho^2}
\left.\frac{\p P}{\p T}\right|_\rho\frac{d\rho}{dr}.
\label{hot}
\eeq
We choose to reduce eqs.~(\ref{con}), (\ref{mom}), and (\ref{hot})
to a set of coupled differential equations for $dv/dr$, $d\rho/dr$, and
$dT/dr$ that make the physics of the wind solution manifest and the solution
to the problem more easily obtained.  We start by eliminating the pressure.
Expanding $P$ differentially in $\rho$ and $T$, we have that
\beq
dP=\left.\frac{\p P}{\p \rho}\right|_T\,\delta\rho+\left.\frac{\p P}{\p T}\right|_\rho \,\delta T.
\label{dp}
\eeq
Defining ($D$) as 
\beq
D=\frac{T}{\rho}\left.\frac{\p P}{\p T}\right|_\rho,
\eeq
we obtain
\beq
c_s^2=c_T^2+\frac{D^2}{C_V\,T}.
\label{csres}
\eeq
where $C_V$ is the specific heat at constant volume,   
$c_s(=\left.\p P/\p \rho\right|_s$) is the adiabatic sound speed,
and $c_T(=\left.\p P/\p \rho\right|_T)$ is the isothermal sound speed.
Taking eq.~(\ref{dp}) and dropping $F_\nu$, we can rewrite eq.~(\ref{mom}) as
\beq
v\frac{dv}{dr}=-\frac{1}{\rho}\left[\left.\frac{\p P}{\p \rho}\right|_T\frac{d\rho}{dr}+
\left.\frac{\p P}{\p T}\right|_\rho 
\frac{dT}{dr}\right]-\frac{GM}{r^2}
\eeq
We can now eliminate $dT/dr$ using eq.~(\ref{hot}) so that
\beq
v\frac{dv}{dr}=-\frac{1}{\rho}\left\{\left.\frac{\p P}{\p \rho}\right|_T\frac{d\rho}{dr}+
\left.\frac{\p P}{\p T}\right|_\rho 
\left[ \frac{T}{C_V\rho^2}
\left.\frac{\p P}{\p T}\right|_\rho\frac{d\rho}{dr}+\frac{\dot{q}}{C_V v}\right]\right\}-\frac{GM}{r^2}
\label{lastv}
\eeq
The terms $d\rho/dr$ can be eliminated using 
eq.~(\ref{con}).  Using eq.~(\ref{csres}) and combining terms proportional to $dv/dr$,
we obtain an expression for $dv/dr$ in terms of thermodynamic quantities returned by the 
equation of state ($D$, $P$, $C_V$, $c_s$, etc.), the basic hydrodynamical 
variables ($\rho$, $v$, and $T$), and the neutrino energy deposition function ($\dot{q}$):
\beq
\frac{dv}{dr}=
\frac{v}{2r}\left(\frac{v_e^2-4c_s^2}{c_s^2-v^2}\right)
+\frac{D}{C_V T}\,\frac{\dot{q}}{c_s^2-v^2},
\label{nwv}
\eeq
where $v_e=(2GM/r)^{1/2}$ is the escape velocity.
Combining eq.~(\ref{nwv}) with eq.~(\ref{con}) and then with  eq.~(\ref{hot}), 
we obtain expressions for $d\rho/dr$ and $dT/dr$, respectively;
\beq
\frac{d\rho}{dr}=\frac{2\rho}{r}\left(\frac{v^2-v_e^2/4}{c_s^2-v^2}\right)
-\frac{\rho}{v}\frac{D}{C_V T}\frac{\dot{q}}{c_s^2-v^2}
\label{nwr}
\eeq
and
\beq
\frac{dT}{dr}=\frac{2}{r}\,\frac{D}{C_V}\left(\frac{v^2-v_e^2/4}{c_s^2-v^2}\right)+
\frac{\dot{q}}{C_V v}\left(\frac{c_T^2-v^2}{c_s^2-v^2}\right).
\label{nwt}
\eeq


\subsection{The General-Relativistic Wind Equations}
\label{section:grequations}

The time-independent hydrodynamical equations for flow in a Schwarzschild spacetime
can be written in the form (Thorne, Flammang, \& Zytkow 1981; Flammang 1982; 
Nobili, Turolla, and Zampieri 1991)
\beq
\frac{1}{vy}\frac{d(vy)}{dr}+\frac{1}{\rho}\frac{d\rho}{dr}+\frac{2}{r}=0,
\label{nobilimass}
\eeq
\beq
\frac{1}{y}\frac{dy}{dr}+\frac{1}{\varepsilon+P}\frac{dP}{dr}=0,
\label{nobilimom}
\eeq
and
\beq
\frac{d\varepsilon}{dr}-\frac{\varepsilon+P}{\rho}\frac{d\rho}{dr}+\rho\frac{\dot{q}}{vy}=0,
\label{nobilien}
\eeq
where $u_r(=vy)$ is the radial component of the fluid four-velocity, 
$v$ is the velocity of the matter measured by a stationary observer,
\beq
y=\left(\frac{1-2GM/rc^2}{1-v^2/c^2}\right)^{1/2},
\eeq
$\varepsilon\,(=\rho c^2+\rho\epsilon)$ is the total mass-energy density, 
$\rho$ is the rest-mass density, 
$P$ is the pressure, $\epsilon$ is the specific internal energy, and 
$\dot{q}$ is the energy deposition rate per unit mass.
These equations assume that the mass of the wind is negligible, 
as in the Newtonian derivation.  Although 
not readily apparent in the form above, eqs.~(\ref{nobilimass})$-$(\ref{nobilien}) 
exhibit a critical point when $v$ equals the local 
speed of sound.  In order to make the solution to this system tractable and 
the critical point manifest we recast the equations in the same form 
as eqs.~(\ref{nwv}-\ref{nwt}):
$$\frac{dv}{dr}=
\frac{v}{2r}\left[\frac{v_e^2}{y^2}\left(\frac{1-c_s^2/c^2}{c_s^2-v^2}\right)
-4c_s^2\left(\frac{1-v^2/c^2}{c_s^2-v^2}\right)\right]$$
\beq
+\frac{D}{C_V T}\,\frac{\dot{q}}{y}\left(\frac{1-v^2/c^2}{c_s^2-v^2}\right),
\label{grv}
\eeq
\beq
\frac{d\rho}{dr}=
\frac{2\rho}{r}\left(\frac{v^2-v_e^2/4y^2}{c_s^2-v^2}\right)
-\frac{\rho}{(vy)}\frac{D}{C_V T}\frac{\dot{q}}{c_s^2-v^2},
\label{grr}
\eeq
and
$$\frac{dT}{dr}=
\frac{2}{r\rho}\frac{D}{C_V}\frac{(P+\varepsilon)}{c^2}
\left(\frac{v^2-v_e^2/4y^2}{c_s^2-v^2}\right)$$
\beq
+\frac{\dot{q}}{C_V (vy)}\left(\frac{(1-D/c^2)c_T^2-v^2}{c_s^2-v^2}\right).
\label{grt}
\eeq
In the above expressions $M$ is the protoneutron star gravitational mass and 
\beq
D=c^2\frac{T}{\varepsilon+P}\left.\frac{\p P}{\p T}\right|_{\rho}.
\eeq
Note that by taking the limits $v/c\ll1$ and $c_s/c\ll1$, we recover the 
Newtonian wind equations in critical form.

\subsection{Conservation and Numerics}

There are two integrals of the flow which we use to gauge the accuracy
of our solution to the Newtonian or general-relativistic wind equations.
The first is the mass outflow rate, obtained from direct integration of
the continuity equation (eq.~\ref{nobilimass}), which yields the eigenvalue
of the steady-state wind problem, $\dot{M}=4\pi r^2 \rho v y$
($y=1$ in the Newtonian limit).
The second is the Bernoulli integral, modified by energy deposition.
In the Newtonian case, 
\beq
\dot{M}\Delta\left(\epsilon+\frac{1}{2}v^2+\frac{P}{\rho}-
\frac{GM}{r}\right)=\int_{R_\nu}^r\,d^3r^{\prime}\,\rho\,\dot{q}=Q(r),
\eeq
where $R_\nu$ is the coordinate radius of the protoneutron star surface
and the $\Delta$ expresses the change in the quantity in parentheses between $R_\nu$ and $r$.
We retain the subscript $\nu$ on $R$ to emphasize that throughout this work
the coordinate radius of the protoneutron star is assumed to coincide
with the radius of decoupling for all neutrino species.
In general relativity, with $\dot{q}=0$, $\gamma h \sqrt{-g_{00}}$ is a constant.  
Here, $\gamma$ is the Lorenz factor and $h$ is the specific enthalpy. 
With a source term, the differential change in neutrino luminosity is given by
\beq
e^{-2\phi}\frac{\p}{\p\mu}(L_\nu e^{2\phi})=-\dot{q},
\eeq
where $d\mu/dr=4\pi r^2\rho \,e^{\Lambda}$.  The expression 
$ds^2=-e^{2\phi}dt^2+e^{2\Lambda}dr^2+r^2d\Omega$
defines the metric.  The total energy deposition rate is then,
\beq
Q=4\pi\int_{R_\nu}^\infty\,dr\,r^2\,\rho\,\dot{q}\,e^{\Lambda}\,e^{2\phi}.
\label{bigQtot}
\eeq

We solve the system of wind equations in critical form using a two-point relaxation
algorithm between $R_\nu$ and the sonic point on an adaptive radial mesh 
(see Thompson, Burrows, \& Meyer 2001; Press et al.~1992; London \& Flannery 1982).
We employ physical boundary conditions for the transonic wind problem, enforcing 
thermal ($\dot{q}=0$) and chemical equilibrium ($dY_e/dr=0$; see \S\ref{section:pnswinds}) 
at $R_\nu$ and $v=c_s$ at the outer edge of the computational domain.  The last 
required boundary condition, which closes the system of equations, is an integral condition
on the electron-neutrino optical depth ($\tau$): $\tau_{\nu_e}(R_\nu)=2/3$.  This condition
is combined with the other two at $R_\nu$ in a triple Newton-Raphson algorithm 
so that all boundary conditions are satisfied simultaneously.  
For modest radial zoning we typically maintain constant $\dot{M}$ and consistent Bernoulli
integral to better than 1\%, having imposed neither as a mathematical constraint 
on the system of wind equations.


\section{Particulars of Protoneutron Star Winds}
\label{section:pnswinds}

\subsection{Electron Fraction Evolution}

The charged-current electron-type neutrino interactions on free nucleons --
$\nu_e n\leftrightarrow e^-p$ and $\bar{\nu}_e p\leftrightarrow e^+n$ -- affect 
the evolution of the electron fraction in radius;
\beq
(vy)\frac{dY_e}{dr}=X_n[\Gamma_{\nu_en}+\Gamma_{e^+n}]-X_p[\Gamma_{\bar{\nu}_ep}+\Gamma_{e^-p}],
\label{yeeq}
\eeq
where $X_n$ and $X_p$ are the neutron and proton fraction, respectively. The $\Gamma$'s are the
number rates for emission and absorption, taken from the approximations of Qian \& Woosley (1996). 
The subscripts denote initial-state particles. Because the resulting nucleosynthesis
in a given wind model is so sensitive to the $Y_e$ at the start of the $r$-process,
we solve the coupled system of the wind equations with $dY_e/dr$ simultaneously.
Ignoring the details of transport and neutrino decoupling near the 
neutrinospheres, the asymptotic electron fraction $Y_e^{\rm a}$ is determined by both 
the luminosity ratio $L_{\bar{\nu}_e}/L_{{\nu}_e}$ and the energy ratio 
$\aveanue/\avenue$, where $\langle\varepsilon_\nu\rangle=\langle 
E_\nu^2\rangle/\langle E_\nu\rangle$, and $E_\nu$ is the neutrino energy.  
To rough approximation (Qian et al.~1993; Qian \& Woosley 1996),
\beq
Y_e^{\rm a}\simeq\frac{\Gamma_{\nu_en}}{\Gamma_{\nu_en}+\Gamma_{\bar{\nu}_ep}}\simeq
\left(1+\frac{\lumanue}{\lumnue}
\frac{\aveanue-2\Delta+1.2\Delta^2/\aveanue}
{\avenue+2\Delta+1.2\Delta^2/\avenue}\right)^{-1},
\label{yeqw}
\eeq
where $\Delta(=m_n-m_p\simeq1.293$ MeV) is the energy threshold
for the $\bar{\nu}_e$ neutrino absorption process, 
$\bar{\nu}_e p \rightarrow n e^+$.
There are several important effects in protoneutron star winds which
act to increase $Y_e^{\rm a}$: (1) The Threshold Effect: even if $\aveanue/\avenue$
is constant in time, if both $\aveanue$ and $\avenue$ decrease, $Y_e^{\rm a}$ must increase as a result of
$\Delta$ and (2) The Alpha Effect: as the flow cools in moving away from
the protoneutron star and $\alpha$-particles are preferentially formed residual 
excess neutrons will capture $\nu_e$ neutrinos, thus increasing
$Y_e^{\rm a}$ (Fuller \& Meyer 1995; McLaughlin, Fuller, \& Wilson 1996).  The $\alpha$-effect 
is particularly important for flows with long dynamical timescales.

\subsection{Neutrino Energy Deposition}

{\bf The Charged Current Processes:} 
In the protoneutron star wind context, the charged-current 
processes ($\nu_e n\leftrightarrow e^-p$ and $\bar{\nu}_e p\leftrightarrow e^+n$)
compete with neutrino-electron/positron scattering as the dominant energy
deposition mechanisms. Ignoring final-state blocking and assuming relativistic 
electrons and positrons, the charged-current specific cooling rate can be written as
\beq	
C_{\rm{cc}}\simeq2.0\times10^{18}\,T^6\,\left[X_p\frac{F_5(\eta_e)}{F_5(0)}+
X_n\frac{F_5(-\eta_e)}{F_5(0)}\right],
\label{ccc}
\eeq
where
$F_n(y)=\int_0^\infty x^n(e^{x-y}+1)^{-1}\,dx$,
$T$ is in MeV, and $\eta_e=\mu_e/T$.  
The heating rate is
\beq
H_{\rm{cc}}\simeq9.3\times10^{18}R_{\nu6}^{\,-2}
\left[\,X_n\, L^{\rm{51}}_{\nu_e}\,\langle\varepsilon^2_{\nu_e}\rangle\,+\,
X_p\, L^{\rm{51}}_{\bar{\nu}_e}\,\langle\varepsilon^2_{\bar{\nu}_e}\rangle\,\right]
\Phi^6\,\Xi(r),
\label{hcc}
\eeq
where $R_{\nu6}$ is the neutrinosphere radius in units of $10^{6}$\,cm,
$L_\nu(=10^{51}L_\nu^{51})$ and $\langle\varepsilon_\nu^2\rangle$ are defined at $R_\nu$, 
$\Phi=[(1-2GM/R_\nu c^2)/(1-2GM/rc^2)]^{1/2}$ is the gravitational redshift, 
and $\Xi(r)$ is the spherical dilution function.
In the vacuum approximation and assuming a sharp neutrinosphere,
\beq
\Xi(r)= 1-\sqrt{1-(R_\nu/r)^2/\Phi^2}.
\label{null}
\eeq
The redshift term, $\Phi$, appearing in eq.~(\ref{null}), accounts for the 
amplification of the heating processes due to the bending of null geodesics in general relativity
(Salmonson \& Wilson 1999).  Cardall \& Fuller (1997) showed that although this
amplification is important, the dominant general-relativistic effect on the
heating rates is due to the gravitational redshift of the
neutrino energy and luminosity (note the $\Phi^6$ in eq.~\ref{hcc}).

With eqs.~(\ref{ccc}) and (\ref{hcc}), the specific energy deposition 
rate (erg\,g$^{-1}$\,s$^{-1}$) due to the charged-current processes 
is simply $\dot{q}_{cc}=H_{cc}-C_{cc}$.

{\bf Inelastic Neutrino  Scattering:}
At high entropies, electron-positron pairs are produced in abundance.
The energy transfer associated with a single neutrino-electron or positron 
scattering event is well approximated by $\omega_i\simeq(\varepsilon_{\nu_i}-4T)/2$, 
where $\omega$ is the energy transfer and $i$ labels the neutrino species (Bahcall 1964).  
Note that $\omega$ allows for both net heating or net cooling, depending
upon the local temperature and the neutrino spectral characteristics.  
A number of researchers have dropped the `$4T$' cooling part from $\omega$ 
(Qian \& Woosley 1996; Otsuki et al.~2000).  This leads to larger net energy
deposition and significant modifications to the wind solution.
We have made a comparison between models with and without this term and 
find that omitting cooling leads to a $40-60$\% increase in $\dot{M}$
and a $10-25$\% decrease in $\tau_{\rm dyn}$.  The asymptotic mechanical luminosity, 
$\dot{M}v^2_\infty/2$, is increased by as much as 80\% and the asymptotic entropy 
is decreased at the 5\% level. The 
largest deviations are in the models with the lowest luminosity and the
highest $s_{\rm a}$.  This is expected because as the luminosity decreases
and the entropy of the flow increases, neutrino-electron scattering contributes
more to the total energy deposition profile.  Because the shape of the
overall heating profile is not significantly affected by omitting the `$4T$' 
cooling term from $\omega$, the asymptotic entropy is only modified slightly.

For inelastic neutrino-electron scattering, the net specific energy deposition rate can be 
approximated by $\dot{q}\simeq cn_e n_{\nu_i}\langle\sigma_{\nu_i e}\,\omega\rangle$,
where $n_e$ and $n_{\nu_i}$ are the number density of 
electrons and neutrinos, respectively, and $\sigma_{\nu_i e}\simeq\kappa_i\,T\,\varepsilon_{\nu_i}$
(Tubbs and Schramm 1975), $\kappa_i=\sigma_o\Lambda_i/2m_e^2$ is a neutrino species 
dependent constant, where $m_e$ is the mass of the
electron in MeV, $\sigma_o\simeq1.71\times10^{-44}$ cm$^2$, and $\Lambda_i$ is
the  appropriate combination of vector and axial-vector coupling constants for neutrino species $i$.
Averaging properly, we find that 
\beqa
\dot{q}_{\nu_i e}&=&
\frac{c}{\rho}\left(\frac{T^3}{(\hbar c)^3}\frac{F_2(\eta_e)}{\pi^2}\right)
\frac{L_\nu}{4\pi r^2 c\langle\varepsilon_\nu\rangle\langle\mu\rangle} 
\Phi^2\,\,\,\,\rm{erg\,\,\,g^{-1}\,\,\,s^{-1}}\nonumber \\
&\times&\left[\frac{\kappa}{2}\langle\varepsilon_\nu\rangle\frac{F_4(\eta_\nu)}{F_3(\eta_\nu)}
T\left(\langle\varepsilon_\nu\rangle\Phi\frac{F_2(\eta_\nu)}
{F_3(\eta_\nu)}-4T\frac{F_3(\eta_\nu)}{F_4(\eta_\nu)}\right)\right],
\label{qdotes}
\eeqa
where $\eta_\nu$ is an effective neutrino degeneracy parameter (Janka and Hillebrandt 1989),
and $\langle\mu\rangle$ is the flux factor, which is related to eq.~(\ref{null})
by $\langle\mu\rangle=R_\nu^2/2\Phi^2\Xi(r)r^2$.
In order to obtain the contribution to the net heating from 
neutrino-positron scattering, $\eta_e\rightarrow-\eta_e$ and one 
must also make appropriate changes to $\Lambda_i$.

{\bf Electron/Positron Annihilation and its Inverse:}
Also at high entropies, cooling and heating due to $e^+e^-\leftrightarrow\nu_i\bar{\nu}_i$ 
must be included. Assuming relativistic electrons and positrons,
and ignoring Pauli blocking in the final state, the cooling rate is 
\beq
C\simeq1.4\times10^{17}T^9\rho_8^{-1}\,f(\eta_e)\,\,\,{\rm ergs\,g^{-1}\,s^{-1}},
\label{coldpr}
\eeq
where $f(\eta_e)=[F_4(\eta_e)F_3(-\eta_e)+F_4(-\eta_e)F_3(\eta_e)]/2F_4(0)F_3(0)$,
$\rho_8$ is the mass density in units of $10^8$ g cm$^{-3}$ and $T$ is in MeV.
The specific heating rate due to the inverse process, 
$\nu\bar{\nu}\rightarrow e^+e^-$, is simply 
(Qian \& Woosley 1996)
\beq
H\simeq1.6\times10^{19}\frac{\Psi(x)}{\rho_8\,R_{\nu\,6}^4}
\Phi^9
\left[\lumanue^{51}\lumnue^{51}(\aveanue+\avenue)+\frac{6}{7}(L^{51}_{{\nu}_\mu})^2\aveunu\right],
\label{heatpr}
\eeq
where $\Psi(x)=(1-x)^4(x^2+4x+5)$, and $x=(1-(R_\nu/r)^2/\Phi^2)^{1/2}$.

\subsection{Equation of State}

At the temperatures and densities encountered in protoneutron star winds,
exterior to the radius of neutrino decoupling, to good approximation, free 
neutrons, protons, and alpha particles may be treated as non-relativistic 
ideal gases. We also include photons and a fully general electron/positron 
equation of state is employed (Evonne Marietta, private communication). 
\cite{sumiyoshi} have found that using a general electron/positron EOS can decrease
the dynamical timescale in the nucleosynthetic region of the wind by as much as a factor of two.
Such a modification is important when considering the 
viability of the neutrino-driven wind as a candidate site
for the $r$-process.  For this reason, a general electron/positron EOS is essential.
Also of importance is the inclusion of alpha particles.  The formation of
alpha particles effectively terminates energy deposition via the processes
$\nu_e n\leftrightarrow e^-p$ and $\bar{\nu}_e p\leftrightarrow e^+n$.
Failure to include alpha particles results in more heating and a broader
energy deposition profile.  The entropy of the flow is thereby higher.
In low luminosity, late-time, high entropy ($s_{\rm a}\sim200$) the
difference in $s_{\rm a}$ is $\sim20$ units.




\section{Results: Spherical Models}
\label{section:results}

For a given $M$, $R_\nu$, and $L_\nu$, the solution to eqs.~(\ref{grv})-(\ref{grt})
yields radial profiles of temperature, density, electron fraction, and velocity.  
From these quantities, one obtains $s_{\rm a}$, $Y_e^{\rm a}$, and $\tau_{\rm dyn}$ --
the critical parameters in determining the neutron-to-seed ratio and, hence,
the resulting $r$-process.
By comparing these numbers with $r$-process nucleosynthesis survey calculations like 
those of Hoffman, Woosley, \& Qian (1997) or Meyer \& Brown (1997), one can quickly 
see if a given wind solution inhabits a point in $s_{\rm a}$--$Y_e^{\rm a}$--$\tau_{\rm dyn}$
space where a robust 3$^{\rm rd}$-peak $r$-process is likely.

In order to map the wind solution space, we have constructed evolutionary tracks from
our steady-state models.  As the supernova commences, we expect the protoneutron
star to have large radius and high neutrino luminosity.  As the cooling epoch proceeds,
the protoneutron star will contract to its final radius (perhaps 10\,km) and the
luminosity may decrease as a power-law or quasi-exponentially in time (Burrows \& Lattimer 1986;
Pons et al.~1999).  The actual time dependence of $L_\nu$ and $R_\nu$ depends sensitively
on the equation of state of dense nuclear matter and the details of transport and
deleptonization of the protoneutron star by neutrinos.

\begin{figure}[t]
\vspace*{0.in}
\hspace*{+.7cm}
\includegraphics[height=10cm,width=10cm]{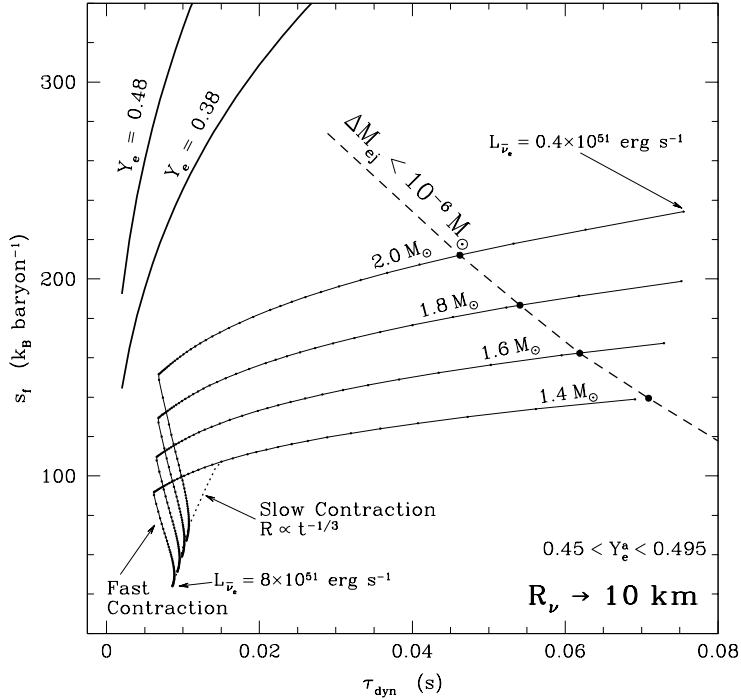}
\vspace*{-.1in}
\caption{Evolutionary tracks of steady-state wind models showing the
various correlations between $s_a$ and $\tau_{\rm dyn}$.  The solid lines
show models for 1.4, 1.6, 1.8, and 2.0\,M$_\odot$ protoneutron stars
(from Thompson, Burrows, \& Meyer 2001).
The thick solid lines show an analytic approximation from
Hoffman, Woosley, \& Qian (1997) for the $s_{\rm a}$ required 
for 3$^{\rm rd}$-peak $r$-process nucleosynthesis for $Y_e^{\rm a}=0.48$ and 0.38.
If $s_{\rm a}$ goes above this line then production of the 3$^{\rm rd}$-peak nuclides is likely.
The thick dashed line denotes the $\tau_{\rm dyn}$ beyond which, for given $M$,
$\dot{M}$ is too small for the wind to contribute significantly to the total galactic $r$-process
budget.
}
\label{fig:fign}
\end{figure}

Figure \ref{fig:fign} summarizes the results of Thompson, Burrows, \& Meyer (2001)
for protoneutron stars with gravitational masses of $1.4$, 1.6, 1.8, and 2.0 M$_\odot$.
Each thin solid line is a sequence of steady-state general-relativistic wind models
in the plane of $\tau_{\rm dyn}$ versus $s_{\rm a}$.  
Each track, for a given $M$, starts with $R\simeq20$\,km and $\lumanue^{51}=8.0$,
corresponding to a total neutrino luminosity of $3.7\times10^{52}$,
where $L_\nu^{51}=L_\nu/10^{51}$.  The models with largest $L_\nu$
and $R_\nu$ have the lowest $s_{\rm a}$ ($\sim50-70$) and moderate $\tau_{\rm dyn}$ ($\sim9$ ms). 
We take $L_\nu\propto t^{-0.9}$ and $R_\nu(t)$ such that the protoneutron
star radius decreases linearly in time from 20\,km to 10\,km in one second
(thin solid lines labeled, `Fast Contraction').  For comparison, we also
include a model that has $R_\nu(t)\propto t^{-1/3}$ for $M=1.4$\,M$_\odot$
(labeled, `Slow Contraction').  In the `Fast' cases, all models move to much
higher $s_{\rm a}$ as $R_\nu$ goes from 20\,km to 10\,km. The $s_{\rm a}$ reached at each $L_\nu$
is set in part by $M/R_\nu(t)$, with the 2.0\,M$_\odot$ model reaching $s_{\rm a}\simeq150$
when $R_\nu$ reaches $10$\,km.  Once the protoneutron star reaches 10\,km, 
$R_\nu$ is fixed and each track makes a sharp turn toward much 
longer $\tau_{\rm dyn}$ and only moderately higher $s_{\rm a}$.  
Due to the relatively slow contraction,
the model with $R_\nu(t)\propto t^{-1/3}$ never exhibits such a sharp 
turn in the $s_{\rm a}$--$\tau_{\rm dyn}$ plane and eventually
joins the `Fast' 1.4\,M$_\odot$ evolutionary track at $\tau_{\rm dyn}\sim0.015$ 
seconds.  Note that after $R_\nu$ has reached 10\,km, the evolutionary tracks
evolve along characteristic curves in the $s_{\rm a}$--$\tau_{\rm dyn}$ plane.
We find that these two $r$-process parameters approximately follow the power law 
\begin{equation}
s_{\rm a}\propto \tau_{\rm dyn}^{0.2}.
\label{sa}
\end{equation}
In their analytic and Newtonian exploration of protoneutron star winds,
Qian \& Woosley (1996) found that $s_{\rm a}\propto \tau_{\rm dyn}^{1/6}$
at constant $R_\nu$ and $M$.  This $1/6^{\rm th}$ scaling and that found
in eq.~(\ref{sa}), from our general-relativistic wind solutions, are to be compared 
with the analytic work of Hoffman, Woosley, \& Qian (1997).  They
find  (see their eqs.~20a \& 20b) that 
the $s_{\rm a}$ required to achieve 3$^{\rm rd}$-peak $r$-process nucleosynthesis
as a function of $\tau_{\rm dyn}$, at constant $Y_e^{\rm a}$, is given by
\begin{equation}
s_{\rm a}\sim730\left(\frac{Y_e^{\rm a}}{0.50}\right)
\left(\frac{\tau_{\rm dyn}}{0.10\,\,{\rm s}}\right)^{1/3}\,\,{\rm k}_{\rm B}\,\,{\rm baryon}^{-1}
\label{eq:hwq}
\end{equation}
for $Y_e^{\rm a}>0.38$. In Fig.~\ref{fig:fign} we show (thick solid
lines) results from eq.~(\ref{eq:hwq}) for  $Y_e^{\rm a}=0.48$ and 0.38.
Because all of the several hundred wind models in Fig.~\ref{fig:fign} have 
$0.45\lesssim Y_e^{\rm a} \lesssim 0.495$, we conclude that models with 
$M=1.4$\,M$_\odot$ fall short of the required entropy by at least a factor of $\sim3$. 
These lines are meant only to delineate the relevant range of $s_{\rm a}$ and
$\tau_{\rm dyn}$ required for 3$^{\rm rd}$-peak nucleosynthesis.  Actual 
nucleosynthetic calculations in the wind profiles themselves are preferred to
the simple comparison on this plot.  Thompson, Burrows, \& Meyer (2001) did just
this, carrying out the full $r$-process calculation in the 1.4\,M$_\odot$ evolutionary track.
They found that nucleosynthesis did not proceed beyond $A\sim100$, in accordance
with the predictions of the survey calculations of Meyer \& Brown (1997) and
Hoffman, Woosley, \& Qian (1997).

That the exponent $1/3$ appears in eq.~(\ref{eq:hwq}) and $0.2$ appears in eq.~(\ref{sa}) 
implies that once $R_\nu$ is set, for a given $M$, the wind cannot evolve
into a region of the $s_{\rm a}$--$\tau_{\rm dyn}$ plane where the $r$-process
can take place.  Put another way, the nearly horizontal lines of constant $R_\nu$,
for a given $M$, in Fig.~\ref{fig:fign} cannot cross the lines of eq.~(\ref{eq:hwq})
at arbitrarily long $\tau_{\rm dyn}$.
Thus, we conclude, along with
Takahashi, Witti, \& Janka (1994), Qian \& Woosley (1996), Sumiyoshi et al.~(2000),
Wanajo et al.~(2000), and Otsuki et al.~(2000),
that winds from protoneutron stars with `canonical' parameters $M=1.4$\,M$_\odot$
and $R_\nu=10$\,km  fail to produce robust $r$-process nucleosynthesis up
to and beyond the 3$^{\rm rd}$ $r$-process peak.  Although the effects
of general relativity are important in determining the dynamical timescales
and entropy of the wind as predicted in Cardall \& Fuller (1997), for reasonable
$Y_e^{\rm a}$'s, the entropy falls short of that required for 3$^{\rm rd}$-peak nucleosynthesis
by a large factor ($\sim3$, slightly better than the factor of $\sim5$ found by
Takahashi, Witti, \& Janka 1994).  


\subsection{Mass Loss}

Having solved for $\dot{M}$ at every point along these evolutionary tracks, and assuming that
$L_\nu(t)\propto t^{-0.9}$, we can calculate the total mass ejected as a function of time:
\beq
M_{\rm ej}(t)=\int_0^t\,\dot{M}(t\pr)\,dt\pr.
\label{mej}
\eeq
Observations of ultra-metal-poor halo stars (see \S\ref{section:observe})
suggest that the astrophysical site for production of the heavy $r$-process nuclides
is universal and acts early in the chemical enrichment history of the galaxy.
The fact that if all supernovae produce this $r$-process signature, then
$10^{-5}-10^{-6}$\,M$_\odot$ of $r$-process material must be ejected per event (Qian 2000)
allows us to constrain the space of relevant wind solutions with eq.~(\ref{mej}).
The heavy dashed line in Fig.~(\ref{fig:fign}), assuming $L_\nu(t)\propto t^{-0.9}$,
shows the point along each evolutionary track in the $s_{\rm a}$--$\tau_{\rm dyn}$ plane
beyond which (to longer $\tau_{\rm dyn}$) $\dot{M}$ is simply too small to generate
the required total mass loss so as to contribute significantly to the total galactic $r$-process
budget.  If $r$-processing begins to the right of this line, less than 10$^{-6}$ M$_\odot$ 
will be ejected. 
Although the position of this $\Delta M_{\rm ej}$
line must change for different $L_\nu(t)$ and $R_\nu(t)$, such a bound must exist
for any cooling model.  
Thus, although the wind may eventually evolve to arbitrarily long dynamical timescales,
we conclude that the range of $\tau_{\rm dyn}$ relevant for $r$-process 
nucleosynthesis is significantly constrained
by consideration of $\dot{M}$ and $M_{\rm ej}$.
For example, the track for $M=1.4$\,M$_\odot$ 
reaches $s_{\rm a}\sim400$ only when $\tau_{\rm dyn}$ is several seconds and $\dot{M}$ 
is of order 10$^{-11}$\,M$_\odot$\,s$^{-1}$.  Hence, even if the $r$-process
could commence in this epoch, it would need to persist for $\sim10^5$ seconds
in order to contribute significantly to the total galactic $r$-process budget.
Conservatively, then, if transonic protoneutron star winds are the primary site for the $r$-process,
this constraint on the amount of mass ejected per supernova implies 
that the epoch of $r$-process nucleosynthesis 
must occur for $\tau_{\rm dyn}$ less than $\sim0.07-0.1$ seconds.

\subsection{Conclusions from Spherical Models}

Models of transonic winds from neutron stars with $M\simeq1.4$\,M$_\odot$ and
$R_\nu\simeq10$\,km fail to produce the heavy $r$-process nuclides.  Models
with much higher gravitational mass and even smaller coordinate radii,
with large neutrino luminosities can achieve 3$^{\rm rd}$-peak nucleosynthesis.
We find that with $M\simeq2.0$\,M$_\odot$ and $R_\nu\simeq9$\,km that 
some 3$^{\rm rd}$-peak nuclides are produced.  It is difficult to understand 
how such massive and compact objects might be created in standard supernova
scenarios.  This has led Thompson, Burrows, \& Meyer (2001) to speculate that
the near environments of collapsars, black holes caused by stellar collapse,
surrounded by a thick accretion disk, might generate outflows much like the wind
solutions described here, but benefiting from the general relativistic effects
as with a $M\simeq2.0$\,M$_\odot$ and $R_\nu\simeq9$\,km protoneutron star.
Barring these possibilities, however, we are left with possible modifications
to the physics described here which might lower $Y_e^{\rm a}$, increase $s_{\rm a}$,
or decrease $\tau_{\rm dyn}$.

Some possibilities for decreasing $Y_e^{\rm a}$ include (1) different $\nu_e$ and
$\bar{\nu}_e$ spectral characteristics (see eq.~\ref{yeqw}), (2) neutrino
oscillations (Qian \& Fuller 1995a,b), and (3) neutrino transport effects.  Of relevance for
(1), changes to the high-density nuclear equation of state may interestingly 
effect the electron and anti-electron neutrino spectra.  Modifications to
the neutrino energy-deposition profile may effect both $s_{\rm a}$ and
$\tau_{\rm dyn}$.  An increase in $\dot{q}(r)$ at fairly large radius 
($50-100$\,km) can increase $s_{\rm a}$ and decrease $\tau_{\rm dyn}$ (Qian \& Woosley 1996).
Such a modification to $\dot{q}(r)$ could be caused by non-standard neutrino
physics or even by magnetic field reconnection (see \S\ref{section:magnetic}).

Finally, one may suspect that the assumption of sphericity is a fundamental problem --
that protoneutron star winds and their ejecta simply cannot be understood fully
in one spatial dimension.  In particular, one may add additional degrees of
freedom and break spherical symmetry, by considering the effects of rotation and magnetic
fields.  As a start to the very complex problem of full magnetohydrodynamic
and general-relativistic outflows, in the following we attempt to quantify some 
of the basic numbers and scalings.


\section{Magnetic Protoneutron Star Winds}
\label{section:magnetic}

The solar wind cannot be explained in detail without consideration of
magnetic effects on the outflow.  This, coupled with the fact that
a class of neutron stars are observed to have very high surface magnetic
field strengths ({\it magnetars}, Kouveliotou et al.~1999; Duncan \& Thompson 1992),
motivates an examination of MHD effects on neutrino-driven protoneutron star winds
and their nucleosynthetic ejecta.

To date, such effects have received little attention.

Qian \& Woosley (1996) speculated qualitatively on the role of magnetic fields 
in their wind solutions, noting that tangled field topologies might
impede the flow in escaping to infinity, but the effects discussed were not quantified.
Nagataki \& Kohri (2001) considered a monopole-like magnetic field 
with rotation in one dimension by restricting their attention to the
equatorial plane.  The formulation was directly analogous to that of 
Weber \& Davis (1967). 
However, because of the complex critical point topology 
encountered in this MHD wind problem, they were unable to assess the importance
of field strengths above $\sim10^{11}$\,G, although they were able to 
consider a variety neutron star rotation periods.  
Recently, 
Cameron (2001) argued qualitatively that core collapse, rotation, and
magnetic fields conspire to form jets and a post-collapse accretion disk
that feeds these jet outflows.

Below we briefly discuss a small subset of the possible effects expected 
from neutrino-driven MHD winds.

\subsection{Non-Spherical Expansion}
\label{section:da}

In a strong magnetic field, the character of the neutrino-driven outflow 
may be significantly modified by the non-spherical divergence of open
field lines, along which the wind is channeled.  Kopp \& Holzer (1976) 
first considered these effects in their models of the solar 
wind in coronal hole regions.  In this case, eq.~(\ref{con1}) becomes
\beq
\frac{d}{ds}(A\rho v)=0,
\label{da}
\eeq
where $ds$ is the differential line element along the magnetic field
and $A(s)$ is an arbitrary area function.  The derivation of equations
analogous to eqs.~(\ref{nwv})-(\ref{nwt}), starting with eq.~(\ref{da}),
is straightforward.  The solution to those equations, however, is complicated
by the possibility that more than one critical point may exist in the
flow if $A(s)$ changes rapidly (Kopp \& Holzer 1976; 
Bailyn, Rosner, \& Tsinganos 1985).  Indeed, standing shocks,
connecting physical solutions, may exist in the flow for rapid areal divergence 
(Habbal \& Tsinganos 1983; Bailyn, Rosner, \& Tsinganos 1985; Leer \& Holzer 1990).
For smoothly and modestly changing $A(s)$, however, the solution 
proceeds as in the spherical case -- the one dimensional problem 
now along $ds$ instead of $dr$. Charboneau \& Hundhausen (1996)
have constructed quasi-two-dimensional models of flow in the field
lines of the open region in a helmet streamer configuration 
(see also Pneuman \& Kopp 1971; Low 1986).  Helmet streamer/coronal hole
magnetic field configurations in the context of the sun have been the
focus of considerable theoretical effort (e.g.~Mestel 1968; Pneuman \& Kopp 1970; Pneuman \& Kopp 1971;
Steinolfson, Suess, \& Wu 1982; Usmanov et al.~2000; Lionello et al.~2002).
In these models, pressure forces, inertia, gravity, and a strong ordered dipole magnetic field
conspire to produce a region of closed magnetic field lines close to the central
star, at latitudes near the magnetic equator.  At the magnetic poles, the flow is
radial.  At intermediate latitudes, between the pole and the closed zone, open magnetic field
lines bend toward the equator close to the star and then extend radially.
In these models $A(s)$ exhibits smooth variations and only for streamlines emerging 
from latitudes very near the closed zone do large deviations from purely radial flow exist
(Charboneau \& Hundhausen 1996).

Taking $A(r)=r^2f(r)$ and $f(r)=(f_{\rm max}\exp[(r-R_1)/\sigma]+f_1)/(\exp[(r-R_1)/\sigma]+1)$,
where $f_1=1-(f_{\rm max}-1)\exp[(R_\nu-R_1)/\sigma]$, as in Kopp \& Holzer (1976), we
have computed several Newtonian wind models for comparison with purely spherical expansion.
This function varies most rapidly near $R_1$, with the change in $f(r)$ over a
radial distance $R_1\pm\sigma$.  With $f_{\rm max}=4$ so that $A(r)$ is four times as 
large at a given radius as spherical expansion and with $R_1=2R_\nu$ and $\sigma=R_\nu$
so that the divergence is smooth, we find that $\tau_{\rm dyn}$ increases by $\sim60$\%,
$\dot{M}$ decreases by $\sim65$\%, $s_{\rm a}$ increases from 68 to 74\,k$_{\rm B}$\,baryon$^{-1}$,
and that the asymptotic mechanical luminosity ($P_{\rm mech}=\dot{M} v_\infty^2/2$) drops by
more than a factor of three.  In contrast, constricting the flow with
$f_{\rm max}=1/4$ yields a much faster wind.  In this case, $\tau_{\rm dyn}$
decreases from 3.2 milliseconds to 1.7 milliseconds.
$s_{\rm a}$ decreases from 68 to 66.5\,k$_{\rm B}$\,baryon$^{-1}$.
$\dot{M}$ and $P_{\rm mech}$ increase by a factor of $\sim3$ and $\sim5.5$, respectively.
Clearly the quality of the areal divergence can significantly influence the
properties of the flow.  Although $s_{\rm a}$ was not affected by more than a few percent,
the changes in $\tau_{\rm dyn}$ evidenced by this simple comparison
imply that a more thorough investigation is warranted. 
We save a detailed exploration of these effects for a future work.

\subsection{Closed Zones \& Trapping}

The ideas of this section have recently been set down in Thompson (2003).
Here, we follow the discussion of Thompson (2003) closely.

\begin{figure}[t]
\vspace*{0.in}
\hspace*{+.5cm}
\includegraphics[height=10cm,width=10cm]{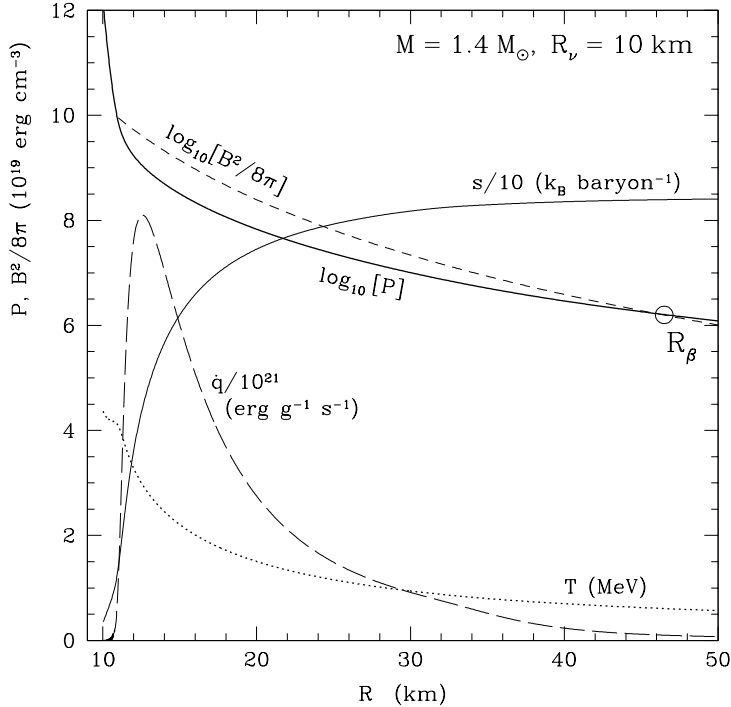}
\vspace*{-.1in}
\caption{Temperature $T$ (MeV, dotted line), entropy $s$ 
(10\,\,k$_{\rm B}$\,\,baryon$^{-1}$, thin solid line), 
energy deposition rate $\dot{q}$ ($10^{21}$\,erg\,g\,s$^{-1}$, long dashed line),
log of the pressure $P$ ($10^{19}$\,\,erg\,\,cm$^{-3}$, thick solid line), and log of the magnetic energy
density $B^2/8\pi$ ($10^{19}$\,\,erg\,\,cm$^{-3}$, short dashed line) for a protoneutron
star wind model with $M=1.4$\,\,M$_\odot$ and $R_\nu=10$\,\,km.  Here, we take
$B=B_0(R_B/r)^3$, where $R_B=11$\,\,km and $B_0=1.5\times10^{15}$\,G.}
\label{fig:all}
\end{figure}

Figure \ref{fig:all} shows profiles of temperature, entropy, energy deposition rate,
and pressure for a high neutrino luminosity protoneutron star wind model
with $M=1.4$\,M$_\odot$ and $R_\nu=10$\,km.  Also shown (thick dotted
line) is the magnetic energy density $B^2/8\pi$.  For simplicity we take 
$B=B_0(R_B/r)^3$, where $R_B$ is a reference radius for the magnetic field
footpoints.  Because of the exponential near-hydrostatic atmosphere in
these wind models (note sharp drop in $P$ in Fig.~\ref{fig:all}), we take 
$R_B=11$\,km.  The surface magnetic field strength $B_0$ is here set to 
$1.5\times10^{15}$\,G.  We define the quantity $\beta=P/(B^2/8\pi)$
and $R_\beta$ as the radius where $\beta=1$.  Here, $R_\beta\sim46.5$\,km.
From this figure it is clear that a $\sim10^{15}$\,G field can dominate
the matter pressure during the wind epoch.  Note that this figure is only for
a single wind model, with a single neutrino luminosity.  As $L_\nu$ drops, 
the pressure profile drops everywhere so that $R_\beta$ moves out in radius for constant $B_0$.
Because $P(r)$ drops everywhere as $L_\nu$ decays, at any instant in time,
a lower $B_0$ is required such that $\beta\sim1$ at some radius.  This 
implies that even though a $10^{15}$\,G field may not dominate the wind dynamics
at early times, if the protoneutron star somehow maintains this field strength
as $L_\nu$ drops, the field will eventually dominate as $P(r)$ decreases
(Thompson 2003).

From this admittedly limited comparison, we conclude that neutrino-driven
winds from protoneutron stars with magnetar-like surface field strengths may 
be significantly affected by the presence of such a field.  

From MHD models and observations of the solar wind, we expect that a
strong magnetic field, that dominates the wind pressure inside $R_\beta$,
may form a closed zone at these radii, at latitudes near the magnetic equator.  
The configuration of the flow would then be analogous to the helmet streamer 
described in \S\ref{section:da} 
(Steinolfson, Suess, \& Wu 1982; Usmanov et al.~2000; Lionello et al.~2002).  
If heating and cooling balance so that $\dot{q}=0$ throughout the closed zone,
this structure may be stable and the matter in this regime permanently
trapped.  However, if heating dominates then the pressure of the trapped matter
will increase.  Thus, if the material is trapped inside $R_\beta$ with $P\ll B^2/8\pi$,
net neutrino energy deposition in the closed zone (see Fig.~\ref{fig:all}) must
increase $P$ to  $\sim B^2/8\pi$.  If this happens, we expect the matter to escape dynamically.
In this way, the closed magnetic field structures that form where $\beta<1$ are unstable.
Importantly for the $r$-process,  because the pressure increase of the matter is caused by 
neutrino heating, it is necessarily accompanied by an increase in the matter entropy
(Thompson 2003).

 
Very roughly, the matter will be trapped for a time set by $P$, $B^2/8\pi$, and $\dot{q}$;
\begin{equation}
\tau_{\rm trap}\sim[B^2/8\pi-P]/[\dot{q}\rho].
\label{tau}
\end{equation}
Assuming that $T$, $\rho$, and $\dot{q}$ do not change 
significantly in $\tau_{\rm trap}$,
the entropy amplification associated with such an increase in pressure is then (Thompson 2003)
\begin{equation}
\Delta s \sim \dot{q}\tau_{\rm trap}/T.
\label{ds}
\end{equation}
When eqs.~(\ref{tau}) and (\ref{ds})  
are evaluated at a characteristic radius for energy deposition (say the half-asymptotic-entropy
radius), they yield an order of magnitude estimate for $\Delta s$. 
For very high $B_0$ and low $P(r)$ (slow winds with low $L_\nu$),
Thompson (2003) found that there is a radius ($R_q$) inside of which cooling balances heating
before $P$ approaches $B^2/8\pi$.  Therefore, in this simple picture, the matter interior
to  $R_q$ is permanently trapped (barring MHD instabilities that might very well arise).
Importantly, for any $B_0$, $R_q$ is always less than $R_\beta$ so that the
trapped matter between these two radii can escape with  $\Delta s$ set by eq.~(\ref{ds}).

\begin{figure}[t]
\vspace*{0.in}
\hspace*{+.5cm}
\includegraphics[height=10cm,width=10cm]{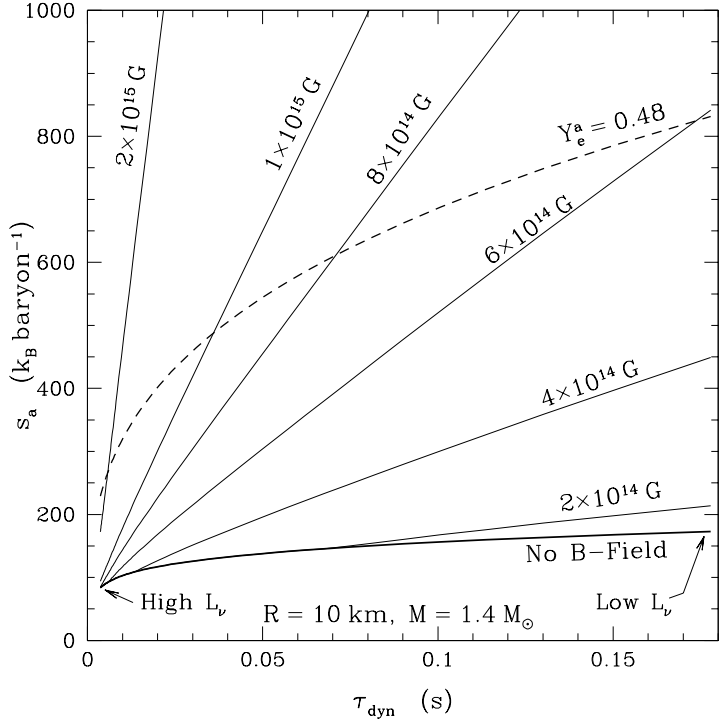}
\vspace*{-.1in}
\caption{Non-magnetic, spherically symmetric wind models (thick solid line)
in the plane of $s_{\rm a}$ versus $\tau_{\rm dyn}$ for protoneutron
star wind models with $M=1.4$\,M$_\odot$ and $R_\nu=10$\,km,
for a large range in neutrino luminosities.  Thin solid lines
show $s_{\rm a}$ as a function of $\tau_{\rm dyn}$, employing
the entropy enhancement as in eq.~(\ref{ds}) for 
$2\times10^{14}\leq B_0\leq2\times10^{15}$\,G.  The
dashed line shows the results of eq.~(\ref{eq:hwq}) for
$Y_e^{\rm a}=0.48$ (from Hoffman, Woosley, \& Qian 1997). 
Figure from Thompson (2003).}
\label{fig:f2}
\end{figure}

Figure \ref{fig:f2} shows $s_{\rm a}$ versus $\tau_{\rm dyn}$ (analogous to Fig.~\ref{fig:fign})
for a large set of protoneutron star wind models (Thompson 2003).  The thick solid line shows spherical,
steady-state models as described in \S\ref{section:results} for constant $R_\nu=10$\,km 
and $M=1.4$\,M$_\odot$.  The dashed line shows
again the analytical results of Hoffman, Woosley, \& Qian (1997) for the $s_{\rm a}$
required, at a given $\tau_{\rm dyn}$, for 3$^{\rm rd}$-peak $r$-process nucleosynthesis
(see eq.~\ref{eq:hwq}; compare with Fig.~\ref{fig:fign}).
Above this dashed line, for $Y_e^{\rm a}=0.48$, the neutron-to-seed ratio
is high enough for a robust $r$-process.  The thin solid lines are
constructed from the non-magnetic wind models (thick solid line), by applying
eqs.~(\ref{tau}) and (\ref{ds}) for $B_0=2\times10^{14}$, $2\times10^{14}$, $4\times10^{14}$,
$6\times10^{14}$, $8\times10^{14}$, $1\times10^{15}$, and $2\times10^{15}$\,G.
From Fig.~\ref{fig:f2} it is clear both that the spherical non-magnetic wind
models fall short of the entropy required for  3$^{\rm rd}$-peak $r$-process
and that for $B_0\sim10^{15}$\,G, the entropy enhancements caused by trapping
may be sufficient to account for this deficit.

There are a number of effects that might decrease the entropy enhancements 
discussed here as a result of trapping in closed magnetic field structures.
Any physical effect that globally disrupts the closed zone on a timescale
much less than $\tau_{\rm trap}$ would significantly undermine the entropy
enhancements estimated in this scenario.  Such effects might include
MHD instabilities, differential rotation, and rapid motion of the magnetic field
footpoints due to convection (Thompson 2003; see also Duncan \& Thompson 1992;
Thompson \& Duncan 1993; Thompson \& Murray 2001).  It is worth noting that 
the very early configuration of the protoneutron star magnetic field is highly 
uncertain and may be a complex of high-order multipoles.  The large-scale
closed zone described in Thompson (2003) might not then obtain.  However, in this
case many closed zones may exist and eq.~(\ref{ds}) may be used to estimate the
entropy enhancement locally at many sites on the surface of the protoneutron
star as closed regions with a variety of $\beta$ are generated and then opened 
by neutrino heating.  If the field is very complex, twisted, or sheared
reconnection may deposit energy in the flow, as first suggested by Qian \& Woosley (1996)
in this context.  Studies of extra energy deposition show that
this may increase or decrease the entropy of the flow, depending crucially
on where the energy is deposited (Qian \& Woosley 1996; Thompson, Burrows, \& Meyer 2001).

\section{Summary, Conclusions, \& Implications}
\label{section:summary}

The subject of protoneutron star winds is relatively young, born only in
the early-1990s.  Much of the physics attending the emergence 
and evolution of these outflows in the just-post-supernova environment is uncertain 
and intimately tied with other outstanding issues in neutron 
star birth: rotation and magnetic fields.  Many of the phenomena
intensively investigated in the context of the sun -- reconnection,
flares, coronal mass ejections, closed loops, prominences, flux
emergence, coronal holes, streamers -- may play important roles 
in determining the nucleosynthetic consequences of the wind/cooling epoch.  

The results presented in \S\ref{section:results} from Thompson, Burrows, \& Meyer (2001)
as well as the results of Takahashi, Witti, \& Janka (1994), 
Qian \& Woosley (1996), Sumiyoshi et al.~(2000), Otsuki et al.~(2000), and Wanajo et al.~(2000)
indicate that spherical steady-state winds from canonical neutron stars 
cannot attain the requisite entropy for robust $r$-process nucleosynthesis.  
Figure \ref{fig:fign} shows, however, that very compact, massive, and luminous
neutron stars may realize a short dynamical timescale, modest entropy $r$-process.
In addition, it may also be that the
actual electron fraction of protoneutron star winds is much lower
than that derived in Thompson, Burrows, \& Meyer (2001) and implied by
the neutron star cooling calculations of Pons et al.~(1999).  If $Y_e^{\rm a}$
could be made to be $\sim0.3$ in the models of Fig.~\ref{fig:fign}, the
1.4\,M$_\odot$ evolutionary track might naturally generate 3$^{\rm rd}$-peak
nuclides.  Then again, perhaps 1.4\,M$_\odot$ neutron stars are responsible
for production of only the $r$-process elements below $A\sim130$ as in 
Fig.~(\ref{fig:sneden}) (see \S\ref{section:observe}).

Section \ref{section:magnetic} shows 
that magnetic effects can change the entropy and dynamical timescale
of a given flow solution considerably for surface field strengths
of order 10$^{15}$\,G.  It may be that only neutron stars born with magnetar-like 
field strengths produce robust $r$-process signatures.  In any case,
the need for multi-dimensional magnetohydrodynamic simulations of wind 
emergence and evolution are required to address many of these still 
open questions more fully.
 

\begin{acknowledgments}
I am indebted to Adam Burrows, Brad Meyer, Eliot Quataert, 
Jon Arons, and Anatoly Spitkovsky for helpful conversations.  
Special thanks go to Evonne Marietta for making her tabular electron/positron EOS
available and to Chris Sneden for providing Figure \ref{fig:sneden}.
Support for this work is provided in part by NASA through a GSRP fellowship
(while the author was at the University of Arizona, Tucson) and by
NASA through Hubble Fellowship
grant \#HST-HF-01157.01-A awarded by the Space Telescope Science
Institute, which is operated by the Association of Universities for Research in Astronomy,
Inc., for NASA, under contract NAS 5-26555.
\end{acknowledgments}


\end{document}